# Origin of discrete electrical switching in chemically heterogeneous vanadium oxide crystals

B. Raju Naik[1], Yadu Chandran[1], Kakunuri Rohini[1], Divya Verma[1], Shriram Ramanathan[2], Viswanath Balakrishnan[1]*

[1]School of Mechanical and Materials Engineering, Indian Institute of Technology, Mandi, Himachal Pradesh-175075, India.

[2]Department of Electrical and Computer Engineering, Rutgers State University, New Jersey, Piscataway, NJ 08854.

*Corresponding author email: viswa@iitmandi.ac.in

**Abstract:**

Electrically driven insulator-metal transitions in prototypical quantum materials such as $VO_2$ offer a foundational platform for designing novel solid-state devices. Tuning the V: O stoichiometry offers a vast electronic phase space with non-trivial collective properties. Here, we report the discovery of discrete threshold switching voltages ($V_{IMT}$) with constant $\Delta V_{IMT}$ between cycles in vanadium oxide crystals. The observed threshold fields over 10000 cycles are ~100X lower than that noted for stoichiometric $VO_2$ and show unique discrete behaviour with constant $\Delta V_{IMT}$. We correlate the observed discrete memristor behaviour with the valence change mechanism and fluctuations in the chemical composition of spatially distributed $VO_2$-$V_nO_{2n-1}$ complex oxide phases. Design of chemical heterogeneity in Mott insulators, therefore, offers an intriguing path to realizing low-energy neuromorphic devices.

Insulator-metal transition (IMT) materials such as $VO_2$ show abrupt changes in electrical conductivity under electrical or thermal stimulus and are of interest as building blocks for artificial neurons[1–3]. Intermediate metastable states might further be of interest for synaptic plasticity[4–7]. Most research on electrical switching characteristics has focused on phase pure compounds of varying strain, dimensionality, W, Al, or oxygen vacancy-doped vanadium dioxide[8–11]. Compositional heterogeneity within a single device channel to optimize the switching voltages represents an early research stage and could lead to reduced threshold switching energies compared to pristine single crystal bulk compounds. It is known that heterophase boundaries play a critical role in creating residual filaments that can help guide conductive filament nucleation[12–16]. Crystalline $VO_2$ is a potential candidate for engineering phase boundaries, mainly because the system can accommodate large vacancy concentrations and host numerous Magnéli-Wadsley phases at different V:O ratios[17–20]. From a material's design perspective, understanding switching mechanisms and electrical properties in heterophase mixtures is essential in optimizing Mott devices. While non-stoichiometry might often be considered a detriment to device function, it is being increasingly examined as a feature in emerging devices, and our results provide a path for reduced switching energies and discrete switching that could be of interest in stochastic circuits.

This article demonstrates a threshold-switching device with discrete $V_{IMT}$ voltages in chemically heterogeneous vanadium oxide single crystals and polycrystalline samples of in-plane and out-of-plane device geometries. Discrete switching at ultra-low voltages in non-stoichiometric vanadium oxide with spatial distribution of $VO_2$ and $V_nO_{2n-1}$ heterophase is caused by electrically driven compositional fluctuations. The number of discrete switching points found in single crystalline samples is significantly higher when compared to the polycrystalline sample. Switching tests conducted over 10,000 I-V cycles show reduced discrete switching points, and eventually, the discrete $V_{IMT}$ voltages are stabilized to a single point. The discrete stochastic switching mechanism was explained by population dynamics of electrochemical species and fluctuations in their valence states under applied electric field stress. This implies that during continuous electric bias switching cycles, the vanadium oxide crystal experiences compositional changes dynamically. Microstructural design of heterogeneity in the V:O system is a promising direction to explore for threshold switching devices with stochastic threshold voltages that are discretized over a large number of cycles. The investigated aspects are also relevant for the fabrication of stoichiometric $VO_2$ thin film devices to enable stable threshold switching over a large number of cycles under optimal unipolar electric fields.

**Results and Discussion**

Figure 1a illustrates the graphical representation of two terminal device on a single crystal $VO_2$ fabricated with an approximate channel length of 40 µm and thickness of 600 nm (Figure S1a-b, Supporting Information). The vapor phase-grown vanadium oxide crystal exhibits distinct dark and bright patterns with random orientations. These contrasting regions are labelled as $VO_2$ (vanadium dioxide) and $V_nO_{2n-1}$ (Magnéli phase vanadium oxide), which signifies the chemical heterogeneity within the crystal. This in-plane device structure has titanium/gold (Ti/Au) as contact electrodes. For electrical characterization, tungsten tips with a tip radius of ~ 7 µm were placed on the Ti/Au electrodes, and voltage was applied across the electrodes while the current passing through the crystal was limited by a set compliance value ($I_{cc}$ = 10 mA). Figure 1b shows representative current versus voltage (I-V) characteristics in linear and semi-log plots (inset). The $V_{IMT}$ (Insulator-Metal Transition) and $V_{MIT}$ (Metal-Insulator Transition) voltages are defined as the specific voltages at which vanadium oxide exhibits a transition in its resistance states, transitioning from a high resistance insulating state to a low-resistance metallic state and, conversely, from metallic to insulating state. Interestingly, multiple current jumps are observed at discrete voltages ($V_{IMT}$), while no considerable variations are seen at $V_{MIT}$. It is known that oxygen deficiencies highly influence the structural stability and IMTs and in turn can reduce the threshold fields significantly. The single crystal device exhibits a large number of discrete IMT switching points (nearly ~9) within the first 100 I-V cycles, with a consistent voltage difference ($\Delta V_{IMT}$) between cycles, which is an intriguing behaviour in $VO_2$. The marked numbers indicate the direction of the current flow in Figure 1b. For precise identification of discrete threshold voltages, all I-V plots are shown in linear scale, with a semi-log plot as inset for visualization. The scatter plot in Figure 1c illustrates the $V_{IMT}$ voltages as a function of the number of cycles, revealing a $\Delta V_{IMT}$ of 0.74 V. In this study, it is observed that from the first initial 100 I-V repetitive cycles, the voltage needed for IMT is significantly smaller compared to stoichiometric $VO_2$. Also, the magnitude of resistance change in the insulator to metal is approximately $10^3$, which is one order less than stoichiometric $VO_2$. We performed material characterizations to investigate the origin of observed discreteness in voltages. Figure 1d shows an optical microscopy image of $VO_2$ single crystals with clear surface contrast in terms of dark and bright regions. We have used yellow and red arrows to illustrate the bright and dark regions within a $VO_2$ single crystal. FESEM image in backscattering mode (inset) confirms the presence of compositional variations in the $VO_2$ single crystals. Further, X-ray photoelectron spectroscopy (XPS) analysis of the elemental composition showed that the single crystal sample comprises 67.6 at. % of +4 oxidation state and 32.3 at. % of +5 oxidation state indicates the mixed phase composition, as shown in Figure 1(e). The X-ray diffraction (XRD) analysis (Figure 1f) shows that the crystals are chemically heterogeneous, as evidenced by a low-intensity impurity peak denoted by '@', which corresponds to Magnéli phase oxide ($V_2O_3$). Detailed structural investigations were carried out by Raman spectroscopy within bright and dark regions (Figure S2a-b, Supporting Information). It has been reported that the lower frequency peaks correspond to V-V vibrations, while the higher frequency peaks (> 300 cm$^{-1}$) correspond to V-O vibrations. Since the $VO_2$ crystals formed using CVD show microdomains with different optical contrasts, we performed Raman spectroscopy in both bright and dark regions. It has been observed that the V-V vibrations at 193 and 225 cm$^{-1}$ have reversed intensity ratios at neighboring bright and dark regions of the same VO$_2$ crystals. Also, the Raman modes of the V-O bond have shifted to lower frequencies. The reversal of intensity ratio of V-V dimers is attributed to the lattice strain present within the sample, affecting the out-of-plane vibrations in the c$_R$ due to the shortening of V-O bonds[21]. Similarly, the shift in V-V and V-O vibrations to lower frequencies also relates to the oxygen deficiency in the sample[22]. Thus, comparing the Raman modes of both bright and dark regions, it can be concluded that the bright region is oxygen deficient compared to the dark region.

In addition, the heterogeneity of the single crystal VO$_2$ was verified and confirmed using various techniques, including energy dispersive spectroscopy (EDS) (Figure S3a-b, Supporting Information), differential scanning calorimetry (DSC) heating and cooling curves (Figure S4a-c, Supporting Information), and *in situ* electrical and mechanical testing (Figure S5a-c, Supporting Information). All these analyses confirm that the as-grown crystal is not fully stoichiometric VO$_2$, instead, it is non-stoichiometric, with majority of volume fraction is dominated by stoichiometric VO$_2$. We conducted similar experiments on polycrystalline samples to verify this unique discrete behaviour. Figure 2a shows the schematic of the electrical measurements conducted in an out-of-plane geometry, with polycrystalline (PC) VO$_2$ sandwiched between a bottom platinum (Pt) electrode and top Ti/Au electrodes. The obtained I-V characteristics for a repetitive cyclic voltage test over 100 cycles display similar conspicuous discreteness in V$_{IMT}$ voltages, as shown in Figure 2b. Thus, the device V$_{IMT}$ voltages are highly unstable and bounce back and forth with multiple I-V cycles. Interestingly, the ΔV$_{IMT}$ between cycles remains constant for the polycrystalline VO$_2$ device. The same is consistent with our observation on single crystal VO$_2$, mentioned earlier. The scatter plot shows a ΔV$_{IMT}$ of 1.02 V over 10000 cycles (Figure 2c). Despite the differences in device geometries, the single and polycrystalline VO$_2$ devices exhibit the same unique discreteness in V$_{IMT}$ with the same equal ΔV$_{IMT}$ behaviour during I-V cycles. The PC-VO$_2$ crystals were further verified by microscopic and spectroscopic analysis. Figure 2d shows the SEM image and indicates the microparticle nature of the PC-VO$_2$. Figure 2e shows the XPS analysis with multiple oxidation states +3, +4, and +5. The XRD analysis shown in Figure 2f confirms that VO$_2$ predominantly dominates the sample. Based on the electrical measurements, it is possible to claim that low-field switching and discrete switching over many cycles are dominated by ionic transport, supported by sweep polarity-dependent electrical I-V measurements. Figure 3a shows I-V characteristics during a positive voltage sweep with multiple IMT switching points. Figure 3b shows stabilized I-V characteristics within 900-1000 cycles, and the scatter plot in Figure 3c shows $V_{IMT}$ distribution over 1000 cycles. Similarly, during the negative sweep, the multiple discrete points quickly get stabilized to a single $V_{IMT}$, and the $V_{IMT}$ distribution over 1000 cycles shows an early stabilization (Figure 3d-f). This provides clear insights into the electroforming of conducting filament with the aid of charge transfer and also reveals the effect of polarity on stochastic threshold switching during voltage sweeps. This suggests that the crystal undergoes compositional fluctuations during the electrical switching cycles especially when the polarity is changed. Dark and bright pattern motion dynamics further verify these compositional changes on single crystal VO$_2$ under microscopy during heating-cooling cycles (Figure S6a-g, Supporting Information). These motion dynamics are observed near room temperature and differ from the typical contrast variations from insulator-metal transition near 68 °C in VO$_2$ crystals (Figure S7a-c, Supporting Information). In earlier works, a systematic reduction in threshold voltage has been demonstrated by introducing oxygen vacancies in VO$_2$ by controlled annealing[23]. Oxygen vacancies have been extensively studied for their effect on electrical transport in VO$_2$. For instance, excess oxygen vacancies increase the carrier concentration and promote metallicity, suppressing abrupt phase transition in VO$_2$[24]. Electrochemical creation and annihilation of oxygen vacancies at the nanoscale by applying an electric field through AFM tip on VO$_2$ is known to affect the transport properties[25]. The approach of engineering the stoichiometry in VO$_2$ by introducing either oxygen vacancies or excess oxygen has been explored for stabilizing rutile or triclinic phases[18,26]. Inherent structural stochasticity with four variants of the M1 phase during insulator–metal transition in $VO_2$ has been observed via *in situ* electrical triggering in TEM. However, it is unclear how the structural stochasticity is directly related to the threshold voltages, especially when a greater number of discrete threshold voltages are observed beyond four structural variatnts[27]. Further, the electric field triggered the formation of $V_5O_9$, conducting filament in $VO_2$, also reported to contribute to non-volatile switching at low threshold voltages[29]. Such non-stoichiometric phases promote resistive switching at very low voltages despite large channel lengths. Earlier reports found that a stoichiometric $VO_2$ demanded significantly high threshold electric fields in the range of $10^6$ to $10^7$ V/m[30]. Conversely, chemically heterogeneous $VO_2$ shows a remarkable reduction of electric fields compared to stochiometric $VO_2$. The estimated threshold electric fields for switching the device vary between $4 \times 10^4$ V/m to $21 \times 10^4$ V/m within the investigated 10000 I-V cycles for $VO_2$ single crystals with a channel length of ~40 µm. Similarly, very low threshold electric fields, varying between $0.6 \times 10^4$ V/m to $2.6 \times 10^4$ V/m, are observed over 10000 repeated I-V cycles in heterogeneous polycrystalline $VO_2$ film with a thickness of 250 nm in the out-of-plane geometry. Note that the estimation of threshold electric fields is carried out with the in-plane channel length of ~40 µm, and in reality, the actual channel length might be much lower as the conducting oxygen deficient Magnéli phase is spatially distributed on the surface of $VO_2$ single crystals.

We correlate the mechanism of low $V_{IMT}$ voltages that show discreteness with constant $\Delta V_{IMT}$ with dynamic fluctuation in composition and related shuttling of oxygen ion vacancies. Figure 4 shows a schematic of the mechanism for discrete threshold voltages during reversible I-V sweeps. Figure 4a and Figure 4b show a representative I-V hysteresis curve with six segments of operation, and crystalline $VO_2$ with dark and bright patterns, respectively. Segment I-III is for a positive voltage sweep and IA-IIIA for a negative voltage sweep. During part I, the reduction process occurs with the creation of oxygen vacancies. As the voltage increases, a conducting filament by $V_nO_{2n-1}$ starts forming by migration of oxygen vacancies across $VO_2$, as shown in Figure 4c. The $V_nO_{2n-1}$ (bright region) filament formation is due to small volume fractions of Magnéli phases between $VO_2$ (dark region) as shown in Figure 4b marked region. This filament formation increases current rapidly, and Joule heating becomes significant to drive the subsequent growth of conducting filament[31]. These filaments act like a guiding path and their formations are sensitive to these dynamic fluctuations in oxygen ion vacancy concentration. In segment II, the insulator-metal transition is triggered once the required threshold voltage is reached, leading to an abrupt increase in current. During segment III, oxidation occurs, and these filaments start rupturing by annihilation of oxygen vacancies by migration of oxygen ions, which results in an abrupt decrease in current (Figure 4d). The above processes are also valid in the negative voltage direction. Since the reduction process and filament formation require more supply of oxygen vacancies over repeated electric cycles, the filament formation path becomes random, generating several $V_{IMT}$. On the contrary, for rupturing, even a little annihilation of oxygen vacancies can break the conducting filament, leading to constant $V_{MIT}$ without much deviations. During repeated electrical cycles, the conducting filament thickness increases, which results in a decrease in $V_{IMT}$, which is evident from the results of both single and polycrystalline $VO_2$ samples.

Nevertheless, the origin of discreteness in $V_{IMT}$ with an equal voltage difference between each consecutive current-voltage characteristic of $VO_2$ needs further validation. It is possible to correlate the electrochemical redox reaction during the electrical triggering and changes in threshold voltages. We use the Nernst equation to understand the mechanism described below.

$$E = E^o - \frac{2.303\,RT}{zF}\log(K) \text{---------(1)}\ ^{32}$$

Where E= Formal potential, $E^o$ = standard potential, K= reaction quotient, z = number of charge transfer, F = Faraday's constant (96500 C mol$^{-1}$) R = gas constant (8.314 J mol$^{-1}$K$^{-1}$) and T= temperature (298K). To make the above expression relevant for gas phase reactions, the reaction quotient has to be expressed in partial pressures of reaction species instead of their ionic concentrations. For all reactions involving the formation of the Magnéli phase ($V_nO_{2n-1}$) in $VO_2$, the expression can be simplified as shown below.

$$E= E^o - \frac{2.303\ RT}{zF} [\frac{1}{2}P(O_2)] \text{---------(2)}$$

Assuming the constant $E^o$ value of 0.34V for all reactions involving reduction of $V^{+4}$ to $V^{+3}$ with fractional charge transfer, formation of $V_2O_3$, $V_3O_5$, $V_4O_7$, $V_5O_9$ etc., values of E, free energy changes ($\Delta G = -nFE$) are calculated and shown in supplementary Table S1. It is possible to evolve several phases derived from a parent structure by crystallographic shear (c.s). Such systematic composition variations have been demonstrated in $TiO_2$ by incorporating c.s structures. For example, Magnéli phases of $Ti_nO_{2n-1}$ wherein n is varied from 4 to 9, resulting in derived structures from $TiO_2$, a rutile type with regular crystallographic shear on (121) planes[33]. Moreover, it is known that slight changes in composition can be accommodated by varying only the orientation without any changes in the number of c.s planes. As mentioned earlier, the recent in situ TEM investigation on VO2 also captured orientation changes within four possible variants during electrical triggering. When the $VO_2$ undergoes slight reduction, the oxygen deficiency may be accommodated by disordered c.s. planes. As the oxygen deficiency increases, they form parallel ordered groups, forming lamellae structures[33]. Eventually they grow in size and form micro domain texture, making the crystal to diphasic heterogeneous system. In the present case, we have observed lamellae-like structures with parallel microdomains of various sizes that are seen in optical and electron microscope images. TEM bright field images and selected area diffraction patterns with superlattice reflections of such micro domain structures in $VO_2$ crystals are shown in Figure S8, Supporting information in consistent with earlier work on $Ti_nO_{2n-1}$.[34] The revealed lamellar and micro domain structures in TEM bright field images, Figure S8(a) and Figure S8(b) along with the superlattice reflections in electron diffraction patterns provide clear evidence for oxygen deficient heterogeneous structure of vanadium oxide.

Interestingly such heterogeneity can be achieved in both oxygen deficient and oxygenated $VO_2$ with excess oxygen. The stoichiometry of crystal can be varied by altering the steps on each c.s planes, keeping the same number of c.s planes. During the reduction and oxidation, creation of excess cation and/or oxygen may flow up or flow down the c.s. planes respectively during voltage sweeps instead of limited only to thermal processing at high temperatures. Note that such minute compositional variations, introduced by annealing at high temperatures with controlled partial pressure of oxygen are known to induced discretized free energy changes in $Ti_nO_{2n-1}$[33]. In the present case, theoretical calculation shows similar feature of discrete free energy changes against their compositions with different n values for $V_nO_{2n-1}$ as shown in Figure 4e. It is known that stability range of each of these Magnéli phases will be small in the order of few kJ/moles. Since the variations in the compositions are minimal, the step heights are also very similar, 0.65 kJ/mole when n varies from 9 to 11. At the same time, for lower values of n, the composition variations and the related step heights are large, evident from Figure 4e. The presented free energy change against the composition for varied ratios of O/V is calculated with the fractional charge transfer, while the mentioned number of cycles is roughly approximated to correlate with the experimental observation of cycle-dependent $V_{IMT}$. Variation in the composition and valence state during repeated cycling influences the value of E and results in discrete free energy changes for forming several non-stoichiometric phases. Because the memristive switching effect in the transition metal oxide memories is often associated with the electrochemical valence change mechanism under electrical triggering, the threshold voltage, $V_T$, and E are correlated[35]. The electrochemical valence state fluctuations may occur during the repeated electrical switching in $VO_2$ in the solid state. Such dynamic electrochemical changes are already reported in the literature for functional oxide-based memristors[36]. The observed discreteness in the $V_{IMT}$ voltages that are seemingly random in their occurrence over 10,000 repeated cycles could be explained by the variation in species undergoing valence change and the formation of several non-stoichiometric Magnéli phases during positive and negative voltage sweeps. In this scenario, the actual triggering voltage will be a random variable $R(V_T)$, which may change due to the valence state reaction's highly repeated threshold voltage, $V_x$ and formal potential (E). Note that any slight variation in the population dynamics of electrochemical species and their valence states between the positive and negative voltage sweeps would affect the $V_T$ as per the following expression with definite voltage difference ($\Delta V_{IMT}$) across multiple switching events.

$R(V_T) = V_x \pm n \Delta V_{IMT}$ -------------------------(3)

Here, the n is an integer (n = 1,2,3, 4, etc) related to the compositional variations in Magnéli phase ($V_nO_{2n-1}$) that forms the conducting channel during the switching cycles. The population dynamics of $V^{+4}$ species undergoing valence state are stochastic but result in switching at discrete threshold voltages dictated by the driving force for electroforming of Magnéli phase channels. Further the $\Delta V_{IMT}$ is related to the step heights in free energy change plot (Figure 4e). The discreteness in threshold voltages with constant $\Delta V_{IMT}$ arises due to the dynamic changes in composition of Magnéli phase wherein 'n' takes larger values above 5. In other words, the mechanism of electrical switching at threshold voltage is governed by insulator-metal transition in $VO_2$; the discrete variation in threshold voltages is arising due to local filament formations and its fluctuations in composition, induced by electrical cycling.

**Conclusion**

Vanadium oxide devices exhibit discrete switching behaviour driven by local fluctuations in composition. Cyclic bipolar switching stability test over 10000 cycles show discrete threshold voltages for several thousand cycles while early stabilization of discrete switching is observed during unipolar switching and confirm the valence change mechanism based memristor performance. The variation in compositional changes and associated free energy changes of Magnéli phases at specific O/V ratios during voltage cycling guides the switching events with equal $\Delta V_{IMT}$. The chemical heterogeneity and the observed fluctuations in fractional valence states resulted in the predictable distribution of discrete $V_{IMT}$ voltages. The demonstrated aspects may provide a novel platform for the practical threshold switching memristor device applications by exploiting the discrete pattern in the switching processes synergistically involving valence change and phase transition.

**Materials and Methods**

The $VO_2$ single crystals were grown by CVD using $V_2O_5$ powder precursor at 900 °C in 8 sccm of Ar gas for 2 hrs. Before introducing the powder precursor in the quartz tube, it was cleaned with Ar purging for 1 hr at 100 sccm of flow rate. Powder precursor was kept inside a quartz boat and placed over a quartz boat on high-temperature brick. Quartz substrate was kept 5 mm away from the powder precursor, and the temperature of the furnace was ramped from 25 °C to 900 °C with a ramp rate of 20 °C /min. The entire growth was carried out in a low-pressure environment with a pressure of $4 \times 10^{-4}$ mbar. For a polycrystalline-$VO_2$, 100 nm platinum (Pt) was deposited on quartz substrate by DC magnetron sputtering, and the same growth conditions of single crystalline $VO_2$ were used with quartz/Pt substrate.

**Device fabrication**

The two terminal planar device geometry on $VO_2$ crystals were fabricated with a commercially purchased shadow mask. The length between two electrodes is 40 µm, and a 50 nm Au was deposited as electrode material using DC magnetron sputtering. For an out-of-plane device with a $VO_2$ memristor, devices were fabricated with a planar device structure. Quartz/Pt/$VO_2$/Au were used as the substrate and electrode materials, respectively. The metal Au electrodes were deposited on $VO_2$ crystals using a shadow mask by DC magnetron sputtering.

**Materials characterization**

The single crystal surface morphology was characterized using a Nikon optical microscope, and thickness was measured with atomic force microscope (AFM) by Bruker. High-resolution images, EDS analysis, and backscattering images were recorded by field emission scanning electron microscopy (FESEM) by Nova Nano SEM-450. Raman vibrational modes were obtained with Raman spectroscopy by Horiba Jobin Yvon at 532 nm laser, crystallographic information using powder X-ray diffraction (XRD) by Rigaku, and elemental analysis using X-ray photoelectron spectroscopy (XPS) by Thermoscientific. The phase transition confirmation of $VO_2$ was done using differential scanning calorimetry (DSC) by NETZSCH Geratebau GmbH. The site-specific nanoindentation was performed by using Hysitron triboindenter 950I. The $VO_2$ crystals grown on a quartz substrate were attached to an iron stub and placed on the magnetic stage. A minimum of 10 indents were performed on each dark and bright region. The details of phase transition confirmation and Raman spectrum in polycrystalline $VO_2$ are provided in Figure S9a-d, Supporting Information. The thickness of films and crystals were measured by atomic force microscopy (AFM) by Bruker Dimension ICON PT. The single crystal structural characterization of $VO_2$ is analysed by high resolution transmission electron microscopy (HR-TEM) FEI-Tecnai G2 20 S-TWIN.

**Device characterization**

Electrical I-V measurements were carried out using Keysight- source measure unit B2902A model at room temperature. Resistance temperature measurements were carried out by connecting a heating stage to the probe station. For all measurements, Ti/Au were used as the electrodes.

**Data availability**

All the data are available within the article and in the supplementary information.

**Author contributions**

B. R. N. carried out the research, starting from the sample preparation, material and device characterization, and involved in the entire investigation and prepared the original manuscript. Y.C. performed nanoindentation, nano-ECR measurements. R. R. carried out CVD growth experiments. Divya Verma recorded optical images of thermally-induced phase transition in $VO_2$ crystals. V. S. R. participated in the discussion and provided critical comments on developing a mechanistic understanding of the work. V. B conceptualized and supervised the overall work and was involved in the original manuscript preparation. All authors have contributed to manuscript writing, reviewing, and editing.

**Competing interests**

The authors declare no competing interests.

**Supporting Information**

Supplementary information is provided.

**Acknowledgements**

This work was supported by the STARS program of the Ministry of Education- India (No. IITM/MHRD-STARS/VB/295). The authors would also like to thank the Advanced Materials Research Centre (AMRC) and centre for design and fabrication of electronic devices C4DFED for providing instrument facilities at IIT Mandi. VB also acknowledges Prof. N. Ravishankar, Indian Institute of Science (IISc), for the helpful discussion.

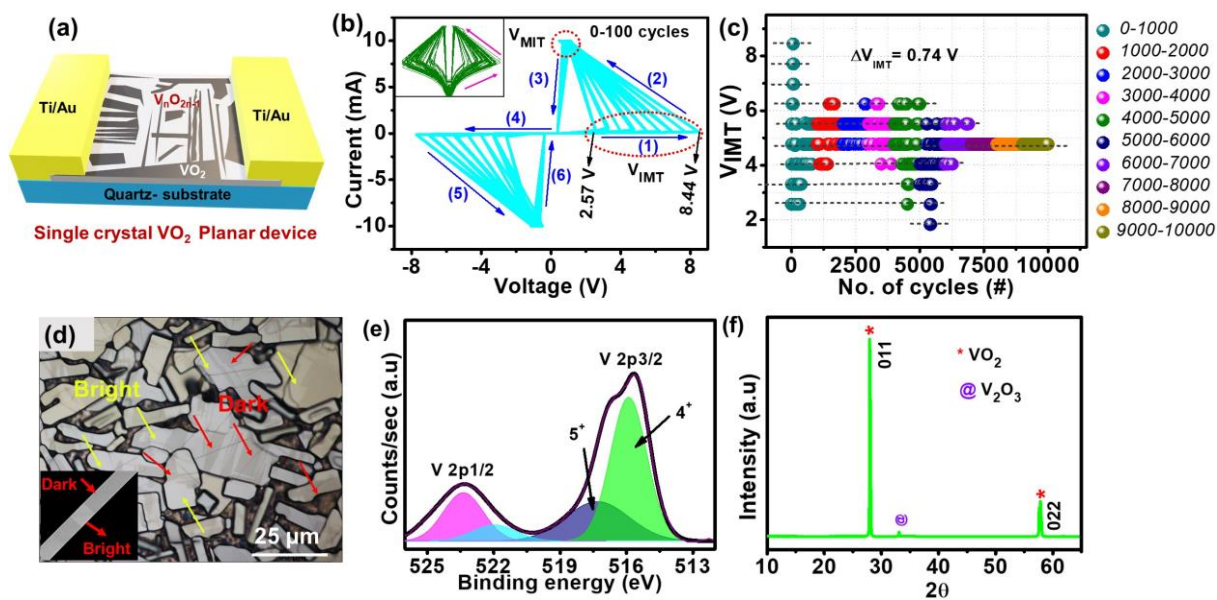

**Figure 1.** Depicts the scheme of the single crystal vanadium oxide device current-voltage measurements and their material characterizations. a) Schematic of chemically heterogeneous single crystal planar device, with an approximate channel length of 40 µm. b) I-V characteristics showing multiple transition points in the linear graph and inset shows semi-log plot. c) IMT scatter plot representing several I-V cycles vs. $V_{IMT}$, having $\Delta V_{IMT}$ of 0.74 V. d) Optical microscopy image of single crystals showing chemical heterogeneity (bright and dark patterns shown by yellow and red arrows) and inset shows the backscattering SEM image. e) V2p scan from XPS analysis representing multiple oxidation states (+4 and +5). f) XRD pattern confirming chemical heterogeneity with presence of Magnéli phase represented with symbol '@'.

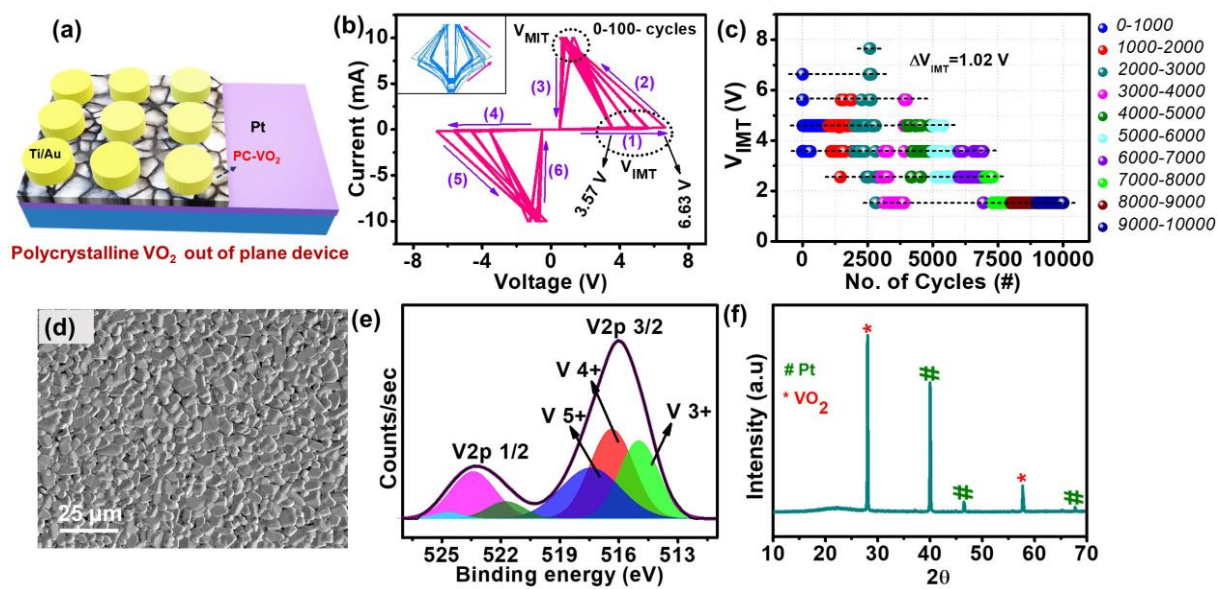

**Figure 2.** Depicts the vertical device current-voltage measurements of the poly crystalline vanadium oxide and their material characterizations. a) Schematic of polycrystalline out of plane device, with platinum and gold as top and bottom electrodes respectively. b) I-V characteristics showing multiple transition points in the linear graph and inset shows semi-log plot. c) IMT scatter plot representing several I-V cycles vs. $V_{IMT}$, $\Delta V_{IMT}$ of 1.02 V. d) SEM image of polycrystalline vanadium oxide showing grain like structures. e) V2p scan from XPS analysis representing multiple oxidation states (+3, +4 and +5). f) XRD pattern confirming the $VO_2$.

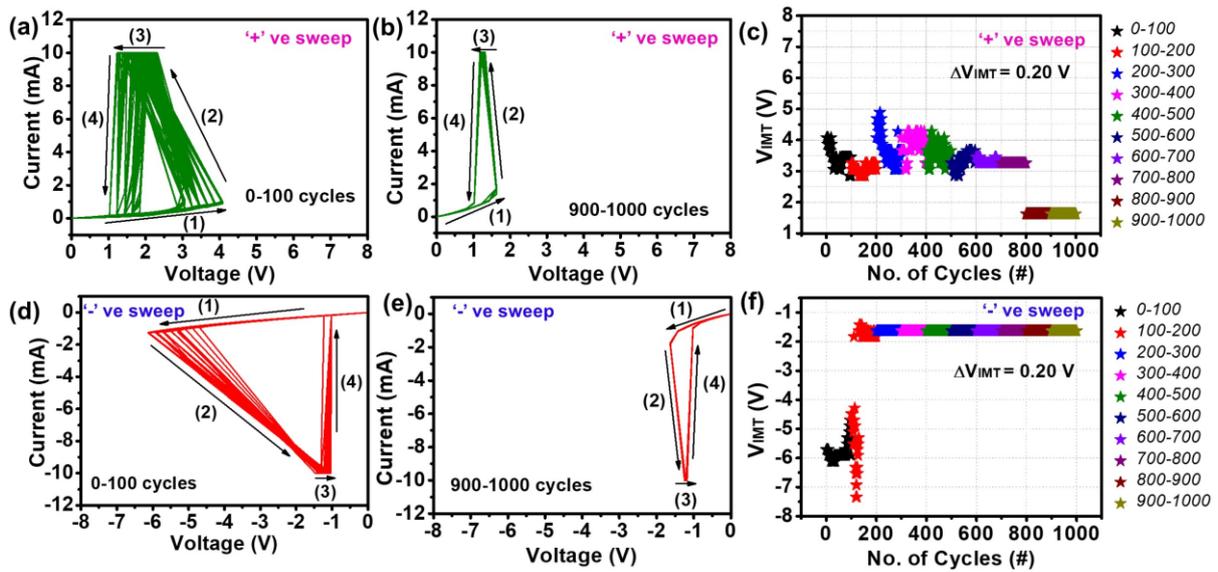

**Figure 3.** Sweep bias polarity dependent electrical transport measurements of single crystal vanadium oxide sample. a) Discrete IMT switching over initial 0-100 I-V cycles during positive sweep. b) Switching between 900-1000 cycles. c) Distribution of $V_{IMT}$ voltages vs number of cycles with $\Delta V_{IMT}$ of 0.20 V. d) Discrete switching in negative sweep for initial 0-100 cycles. e) Switching between 900-1000 cycles. f) Distribution of $V_{IMT}$ voltages vs number of cycles with $\Delta V_{IMT}$ of 0.20 V.

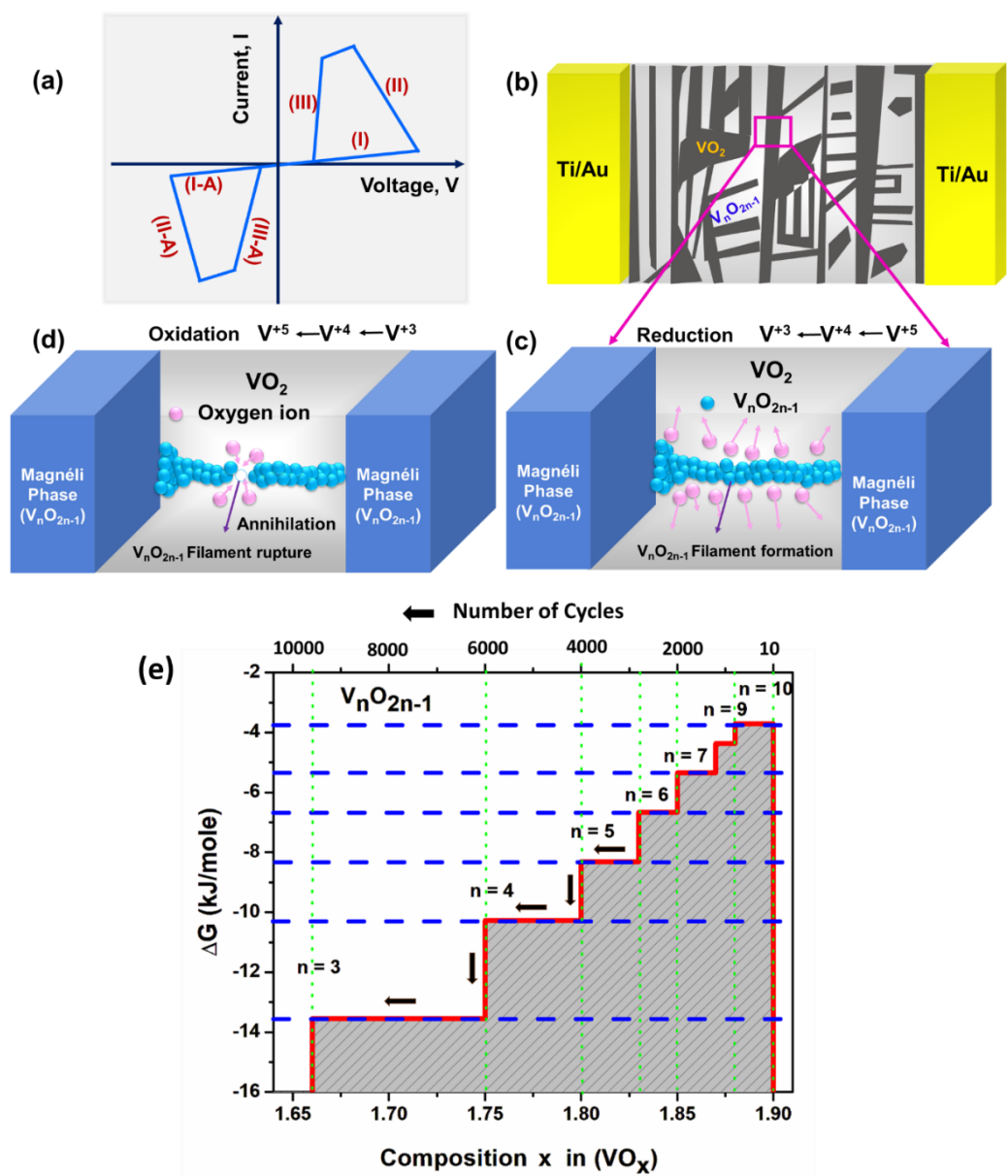

**Figure 4.** Schematic illustration of the conduction mechanism of chemically heterogeneous vanadium oxide with the help of oxygen vacancies and ionic movement. a) Schematic of I-V hysteresis. b) Schematic representation of chemically heterogeneous vanadium oxide device with bright and dark patterns. c) I and I-A represent the vacancy-induced filament formation across Magnéli phases, and II and II-A represent the phase change of $VO_2$. d) III and III-A represent the filament rupturing due to vacancy annihilation. (e) Discrete free energy changes against compositional evolution are observed during electrical switching cycles. The discrete free energy changes with respect to the composition of complex oxides are calculated for redox reactions involving fractional charge transfers, while the mentioned number of cycles is approximate based on experimental observations.